\documentclass[12pt,reqno]{amsart}
\usepackage{hyperref}
\usepackage{appendix}
\usepackage{graphicx}
\usepackage{amscd}
\usepackage{amsmath}
\usepackage{epsfig}
\usepackage{amsfonts}
\usepackage{amssymb}
\usepackage{setspace}
\usepackage{listings}
\usepackage{framed}
\usepackage{xcolor}

\colorlet{shadecolor}{gray!20}
\providecommand{\U}[1]{\protect\rule{.1in}{.1in}}
\providecommand{\U}[1]{\protect\rule{.1in}{.1in}}
\allowdisplaybreaks

\theoremstyle{plain}

\numberwithin{equation}{section}

\input{tcilatex}
\newcommand{\grayitss}[1]{\textcolor{gray}{\textsf{\textit{#1}}}}
\lstdefinestyle{matitalic}{
  language=Mathematica,
  mathescape=true,
  basicstyle=\ttfamily\small,
  columns=fullflexible,
  keepspaces=true,
  breaklines=true,
  upquote=true,
  literate=
    {In[}{{\grayitss{In[}}}1
    {]:=}{{\grayitss{]:=}}}1
    {Out[}{{\grayitss{Out[}}}1
    {]=}{{\grayitss{]=}}}1,
 }
\lstnewenvironment{mat}{\lstset{style=matitalic}}{}

\begin{document}
\title[Oganesson vs Uranium Hydrogen-like Ions and Beyond]{Oganesson versus Uranium
Hydrogen-like Ions and Beyond \\
(from the Viewpoint of Old Quantum Mechanics)}
\author{Kamal Barley}
\address{Department of Mathematics, College of Arts and Sciences, Howard
University, 2141 6th St NW, Washington, DC 200059, U.S.A.}
\email{Kamal.Barley@howard.edu}
\author{Andreas Ruffing}
\address{Landeshauptstadt M\"{u}nchen, Seminar TMG Universit\"{a}t, Referat f%
\"{u}r Bildung und Sport, Marienplatz~8, 80331 M\"{u}nchen, Germany}
\email{alruffing@web.de}
\author{Sergei K. Suslov}
\address{School of Mathematical and Statistical Sciences, Arizona State
University, Tempe, AZ 85287--1804, U.S.A.}
\email{sks@asu.edu}
\date{September 23, 2025}

\begin{abstract}
We compare, within the framework of the old Bohr-Sommerfeld atomic model, Uranium $(Z=92)\/$ versus
hypothetical Oganesson $(Z=118)\/$ relativistic hydrogen-like ions and beyond. 
Existence of self-intersecting orbits in the super strong static Coulomb field of superheavy transuranium elements,
including the Oganesson,
is demonstrated with the aid of Mathematica computer algebra system. 
A possibility of a similar `Oganesson-type' effect in a strong gravitational field is also mentioned.
Appearance of a multi self-intersecting trajectory is demonstrated
in a hypothetical Unbibium, an element with Z = 122 and beyond with $Z\le135$.
A topological classification of the strength for Coulomb fields in the superheavy transuranium elements may be given in terms of the `winding' numbers.
\end{abstract}

\dedicatory{\begin{center}
{\scriptsize{Dedicated to the 100th anniversary of the birth of wave mechanics \cite{SchrQMI}.}}
\end{center}} 
\maketitle

\noindent
{\scriptsize{Hence it was necessary to carry out the calculation as far as
possible according to the method of the older quantum theory; the
inferences drawn can then later be taken over directly into wave mechanics.%
}

\begin{flushright}
\it{Arnold Sommerfeld} \cite[p.~258]{SomAS} 
\end{flushright}
}

\bigskip

\section{Transuranics, Fine Structure Formula, and Orbits\/}

Transuranic elements are chemical elements with atomic numbers greater than $92$ (the atomic number of uranium). All of them are unstable and radioactive, and most must be synthesized in a laboratory \cite{Kragh18}.
The high positive charge of the nucleus creates an extremely strong electric field that affects the behavior of all orbiting electrons.
For the innermost electrons, the electrostatic attraction is so strong that their velocities approach the speed of light. 
The Bohr-Sommerfeld model was an early attempt to explain atomic structure by incorporating elliptical orbits and special relativity into the Bohr model \cite{Barleyetal2025, Kragh2012, KraghBohr}. However, its simplified, semiclassical approach has fundamental limitations that prevent it from accurately describing transuranic elements.
Nonetheless, we would like to explore this ``missing opportunity" for highly charged superheavy one-electron systems.

The original Sommerfeld fine structure formula is given by \cite{Somm1916, SomAS}:
\begin{equation}
\frac{E_{n_{r},n_{\theta }}}{mc^{2}}=\left[ 1+\frac{\alpha ^{2}Z^{2}}{\left(
n_{r}+\left( n_{\theta }^{2}-\alpha ^{2}Z^{2}\right) ^{1/2}\right) ^{2}}%
\right] ^{-1/2} \qquad (Z<137), \label{BidSomEnd}
\end{equation}%
where $n_{r}$ (the {\textit{radial quantum number}}) and $n_{\theta }$ (the {%
\textit{azimuthal quantum number}}) are positive integers. This result made
it possible to explain, for the first time, the fine structure of spectral
lines. (For further details, see \cite{Barleyetal2021, Barleyetal2025, Elyashevich1985, KraghBohr, Reed}
and the references therein.)

In standard geometrical terms, the orbit equation is%
\begin{equation}
\frac{1}{r}=\frac{1+\epsilon \cos \left( \omega \theta \right) }{a\left(
1-\epsilon ^{2}\right) }.  \label{BidSom9}
\end{equation}%
With the eccentricity $\epsilon $, we have, for $\omega \theta = \phi =0$, the
perihelion distance $r_{\text{min}}=a(1-\epsilon )$, and for $\omega \theta = \phi =\pi $,
the aphelion distance $r_{\text{max}}=a(1+\epsilon )$.

As stated in original works \cite{Biedenharn1983, LaLif2, SomAS}, classical
relativistic Kepler orbits have the form of conic sections as in the
nonrelativistic case, but with a new angular variable $\phi =\omega \theta .$
Thus, for elliptical orbits (bound states), the motion from one perihelion ($%
\phi =0$) to the next ($\phi =2\pi $) requires $\theta =2\pi /\omega \,$,
with a per-revolution shift of $\Delta \theta =2\pi /\omega -2\pi .$%
\footnote{In the terminology of classical work \cite{SomAS} where only sufficiently small $\Delta \theta $ are discussed (Figure~\ref{Figure1}).
Throughout this article we shall keep the original name, `elliptical' curve, although for certain $\omega$'s a different topological structure may occur such as orbits with different `winding' numbers and with multiple points of self-intersections (see Figures~\ref{Figure2} to \ref{Figure7} below). 
}

Quantized values of parameters of the electron's elliptical orbits (\ref%
{BidSom9}) are evaluated in \cite{Barleyetal2025} as follows%
\begin{eqnarray}
\omega _{n_{\theta }}n_{\theta } &=&\left( n_{\theta }^{2}-\alpha
^{2}Z^{2}\right) ^{1/2},  \label{BidSom18} \\
\epsilon _{n_{r},n_{\theta }} &=&\sqrt{n_{r}}\cdot \frac{\left( n_{r}+2\sqrt{%
n_{\theta }^{2}-\alpha ^{2}Z^{2}}\right) ^{1/2}}{n_{r}+\sqrt{n_{\theta
}^{2}-\alpha ^{2}Z^{2}}},  \label{BidSom19} \\
a_{n_{r},n_{\theta }} &=&\frac{a_{0}}{Z}\left( n_{r}+\sqrt{n_{\theta
}^{2}-\alpha ^{2}Z^{2}}\right)  \label{BidSom20} \\
&&\times \sqrt{\alpha ^{2}Z^{2}+\left( n_{r}+\sqrt{n_{\theta }^{2}-\alpha
^{2}Z^{2}}\right) ^{2}} ,  \notag
\end{eqnarray}%
where $a_{0}=\hbar ^{2}/(me^{2})$ is the familiar Bohr radius and $\alpha =
e^{2} / (\hbar c)$ is the fine structure formula. These formulas generalize
the circular orbits.\footnote{%
Classical solutions of the relativistic Kepler problem are also discussed in 
\cite[pp.~481--482]{Goldstein} and \cite[pp.~100--102]{LaLif2}.}

\section{Some Trajectories of Finite Motion: Uranium vs Oganesson and Beyond\/}

We believe that utilizing the Bohr–Sommerfeld model to contrast $U^{91+}$, $Og^{117+}$, hypothetical $Ubb^{121+}$ hydrogen-like ions, and beyond gives a fresh pedagogical and historical perspective -- it links early quantum theory \cite{Somm1916, SomAS} to modern superheavy transuranium relativistic systems \cite{Oganessian2017} in the quasi-classical approximation (using the Dirac equation as a next step \cite{Barleyetal2025}).

\subsection{Uranium\/}
Uranium (\textit{U}) is a naturally occurring, radioactive, heavy metal with
atomic number $Z=92\/$, known for its ability to release concentrated energy
through nuclear fission, which powers nuclear reactors and weapons. Quantum
electrodynamics of the hydrogen-like Uranium ion had been studied in \cite%
{Gum2005,Gum2007} (see also \cite{FleKar1971, Mohretal1998, Oganessian2017, ShabQED, ShabQEDHI, SusRelInt} and references therein).
We use Mathematica for the following animations \cite{BarleySusMathTwo}.

\begin{figure}[hbt!]
\centering
\includegraphics[width=0.557\textwidth]{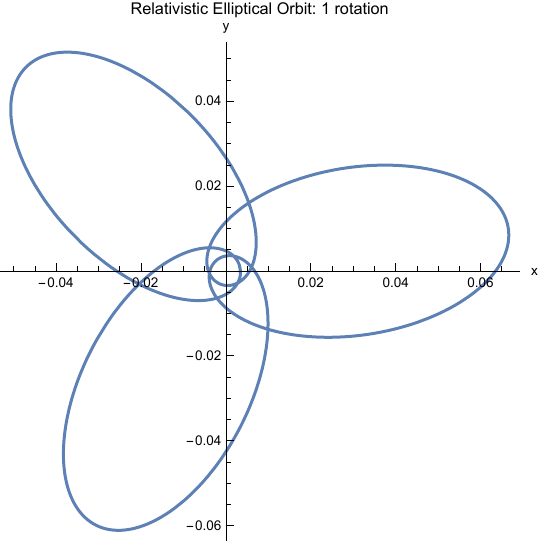}
\caption{Kepler's Classical Elliptical Curve Counterclockwise Motion in Relativistic Uranium Hydrogen-like
Ion $U^{91+}\/.$ The `winding' number is one.}
\label{Figure1}
\end{figure}

In the Bohr-Sommerfeld atomic model under consideration, even for the strong electric field of the Uranium nucleus, 
a single electron stays on classical relativistic elliptical orbits with advancing perihelion, 
in a form of open ``rosettes" \cite{Biedenharn1983, LaLif2, SomAS} (see Figure~\ref{Figure1} with the standard orbit `winding' number one).  
%
%

%
In the above animation of the relativistic Kepler motion of an electron in
hydrogen-like Uranium ion, $U^{91+}$, when $Z=92$, the nucleus is situated in
the fixed focus at the origin. For this quantum state, we choose $%
n_{r}=n_{\theta }=1$ in the fine structure formula (\ref{BidSomEnd}). By (%
\ref{BidSom18})--(\ref{BidSom20}), one gets: $\omega =0.741\ 135$, $\epsilon
=0.904\ 882$, and $a=0.035\ 3163$, in Bohr's atomic units. The perihelion
and aphelion move along two concentric circles around the nucleus with
radii: $r_{\text{min}}=a(1-\epsilon )=0.003\ 3592\ 09$ and $r_{\text{max}%
}=a(1+\epsilon )=0.067\ 2734\ 72$, respectively. In the animation, the
geometrical loci of the successive perihelia and aphelia, the outer and
inner envelopes of the orbit \cite{SomAS}, are not shown for simplicity, Figure~\ref{Figure1}. 
Counterclockwise rotation of the ellipse over one period is given by 
$\Delta \theta =2\pi \left( 1/\omega -1\right) =$ $2.194\ 61.$
%

%
\begin{figure}[tbh]
\centering
\includegraphics[width=0.557\textwidth]{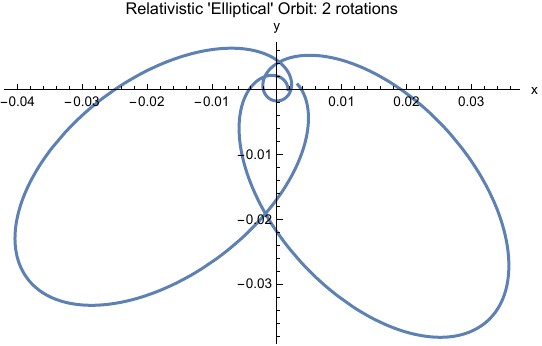}
\caption{Kepler's `Elliptical' Curve Clockwise Motion with Self-intersection in
Hypothetical Copernicium Hydrogen-like Ion $Cn^{111+}\/.$ The `winding' number is two.}
\label{Figure2}
\end{figure}

\vfill\eject\newpage

\subsection{Copernicium}  
Copernicium (\textit{Cn}) is a superheavy, synthetic, and radioactive chemical element with atomic number $Z=112$, discovered in 1996 at the Gesellschaft f\"{u}r Schwerionenforschung (GSI) in Darmstadt, Germany.
It was officially named after the astronomer Nicolaus Copernicus in 2010 \cite{Kragh18}.

Our Mathematica animations show that in a super strong electric field of the Copernicium nucleus, 
the classical relativistic finite Kepler orbits of a single electron may change its topology, 
see Figure~\ref{Figure2} with the orbit `winding' number two%
\footnote{The so-called ``chemical orbit'' or ``double necklace'' in terminology of the American cosmologist John
Archibald Wheeler \cite{CeulemansThyssen2018, Powers1971, Wheeler1971}.
}.
In the above animation of the relativistic Kepler motion of an electron in
hydrogen-like Copernicium ion, $Cn^{111+}$, when $Z=112$, the nucleus is
situated in the fixed focus at the origin. For this quantum state, we choose
$n_{r}=n_{\theta }=1$ in the fine structure formula (\ref{BidSomEnd}). By (%
\ref{BidSom18})--(\ref{BidSom20}), one gets: $\omega =0.576\ 207$, $\epsilon
=0.930\ 786$, and $a=0.024\ 9872$, in Bohr's atomic units. The perihelion and
aphelion move along two concentric circles around the nucleus with radii: $%
r_{\text{min}}=a(1-\epsilon )=0.001\ 7294\ 7$ and $r_{\text{max}}=a(1+\epsilon
)=0.048\ 2449$, respectively. In the animation, the geometrical loci of the
successive perihelia and aphelia, the outer and inner envelopes of the orbit
are not shown for simplicity. Total rotation of the self-intersecting
`ellipse' over one period is given by $\Delta \theta =2\pi \left( 1/\omega
-1\right) =4.621\ 2.$ 

At the point of self-intersection one gets,
approximately with the help of Mathematica, that
$r(2.379\ 9243\ 5238\ 77007)=r(8.524\ 4563\ 1673\ 413)=0.002\ 8192\ 8$
with 
$r(8.524\ 4563\ 1673\ 413) - r(2.379\ 9243\ 5238\ 77007)= -1.734\ 72 × 10^{-18} $
and
$\omega \times (8.524\ 4563\ 1673\ 413 + 2.379\ 9243\ 5238\ 77007)/2\\ = 3.141\ 59 = \pi\/.$

\subsection{Oganesson\/} 
Oganesson (\textit{Og}) is the heaviest synthetic element on the periodic
table, with atomic number $Z=118\/$. Discovered by scientists at the Joint
Institute for Nuclear Research in Russia and Lawrence Livermore National
Laboratory in $2006\/$ \cite{Oganessian2017}, it is named after physicist Yuri Oganessian (see \cite%
{Chap2018, Colloquium2019, Kaygorodovetal2021, Kragh18, Oganessian2006, Oganessian2012} and references therein). 

\begin{figure}[tbh]
\centering
\includegraphics[width=0.557\textwidth]{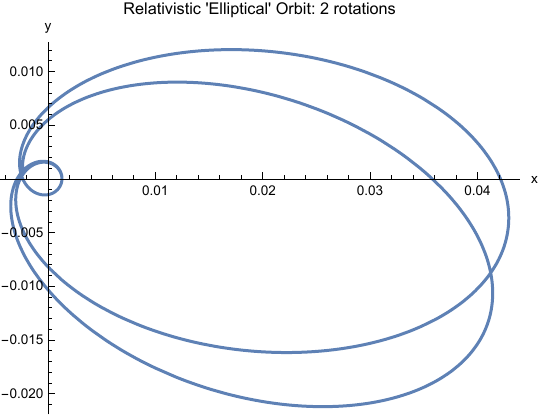}
\caption{Kepler's `Elliptical' Curve Clockwise Motion with Self-intersection in
Hypothetical Oganesson Hydrogen-like Ion $Og^{117+}\/.$ The `winding' number is two.}
\label{Figure3}
\end{figure}

Our Mathematica animations show that in a super strong electric field of the  Oganesson nucleus, 
the classical relativistic finite Kepler orbits of a single electron may change its topology, 
see Figure~\ref{Figure3} with the orbit `winding' number two%
%

%

%
In the above animation of the relativistic Kepler motion of an electron in
hydrogen-like Oganesson ion, $Og^{117+}$, when $Z=118$, the nucleus is
situated in the fixed focus at the origin. For this quantum state, we choose
$n_{r}=n_{\theta }=1$ in the fine structure formula (\ref{BidSomEnd}). By (%
\ref{BidSom18})--(\ref{BidSom20}), one gets: $\omega =0.508\ 457$, $\epsilon
=0.941\ 479$, and $a=0.022\ 2041$, in Bohr's atomic units. The perihelion and
aphelion move along two concentric circles around the nucleus with radii: $%
r_{\text{min}}=a(1-\epsilon )=0.001\ 2994$ and $r_{\text{max}}=a(1+\epsilon
)=0.043\ 1087$, respectively. In the animation, the geometrical loci of the
successive perihelia and aphelia, the outer and inner envelopes of the orbit
are not shown for simplicity. Total rotation of the self-intersecting
`ellipse' over one period is given by $\Delta \theta =2\pi \left( 1/\omega
-1\right) =6.074\ 18.$ 
At the point of self-intersection one gets,
approximately with the help of Mathematica, that
$r(3.074\ 0374\ 2053\ 6865)=r(9.283\ 3297\ 0962\ 4294)=0.002\ 5044$
with 
$r(9.283\ 3297\ 0962\ 4294) - r(3.074\ 0374\ 2053\ 6865)= 4.33\ 681 × 10^{-19} $
and
$\omega \times (9.283\ 3297\ 0962\ 4294 + 3.074\ 0374\ 2053\ 6865)/2 = 3.141\ 5926\ 5358\ 9793 = \pi\/.$
%

%
\begin{figure}[tbh]
\centering
\includegraphics[width=0.557\textwidth]{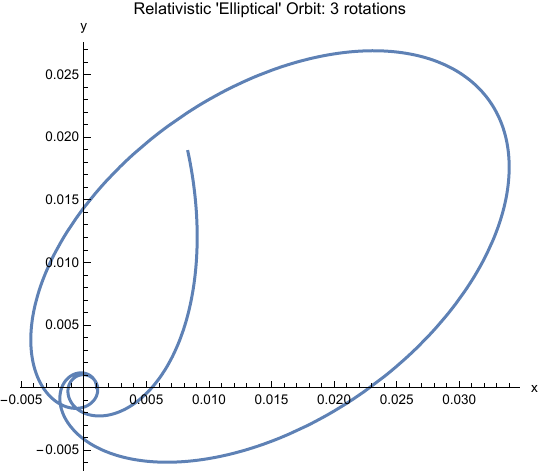}
\caption{Kepler's `Elliptical' Counterclockwise Motion with Four Self-intersections in
Hypothetical Unbibium Hydrogen-like Ion $Ubb^{121+}\/.$ The `winding' number is three.}
\label{Figure4}
\end{figure}
%

\subsection{Unbibium\/} 
Element $122$, is a hypothetical chemical element with the placeholder symbol $Ubb$. 
In the above animation of the relativistic Kepler motion of an electron in a 
hydrogen-like Unbibium ion, $Ubb^{121+}$, the atomic number $Z=122$, the nucleus is situated in the
fixed focus at the origin. For this quantum state, we choose $n_{r}=n_{\theta}=1$ 
in the fine structure formula (\ref{BidSomEnd}). By (\ref{BidSom18})--(\ref{BidSom20}), one gets:
$\omega=0.455\ 419$, $\epsilon=0.949\ 782$, and $a=0.020\ 3534$, in Bohr's atomic units.

The perihelion and aphelion move along two concentric circles around the
nucleus with radii: $r_{\text{min}}=a(1-\epsilon)=0.001\ 02211$ and
$r_{\text{max}}=a(1+\epsilon)=0.039\ 6847$, respectively. In the animation,
the geometrical loci of the successive perihelia and aphelia, the outer and
inner envelopes of the orbit, see Figure~\ref{Figure4}, are not shown for simplicity.
Rotation of the ellipse over one period is given by $\Delta\theta
=2\pi(1/\omega-1)=7.513\ 3.\/$ The `winding' number is three.

At the point of the first self-intersection one gets, approximately,
$r(10.002448573515926)=r(3.7940322175405243)=0.00234067$ with
$r(10.002448573515926)-r(3.7940322175405243)=0.\/$ Here, $\omega
(10.002448573515926+3.794032217540524)/2=3.14159=\pi.\/$

At the second self-intersection, approximately,
$r(13.18\ 1425\ 7027\ 0781)=r(0.615\ 0550\ 8834\ 86369)$
$=0.001\ 04189$ and 
$r(13.18\ 1425\ 7027\ 0781) - r(0.615\ 0550\ 8834\ 86369)=4.336\ 81\times 10^{-19}\/.$
Here, $\omega(13.18\ 1425\ 7027\ 0781 + 0.615\ 0550\ 8834\ 86369)/2=3.141\ 59=\pi\/.$

We have two more consecutive self-intersections:
$r(16.93\ 2961\ 5821\ 129)=r(10.66)=0.001\ 7562$ with
$r(16.93\ 2961\ 5821\ 129)-r(10.66\ 0000\ 0000\ 0000\ 2)=-8.67362\times
10^{-19}.$ Here, $\omega(16.93\ 2961\ 5821\ $
$129+10.66\ )/2=6.28319=2\pi.$

And, 
$r(20.07\ 9666\ 0982\ 3604)=r(7.513\ 2954\ 8387\ 6864)=0.022\ 845$ with
$r(20.07\ 9666\ 0982\ 3604)$ $-r(7.513\ 2954\ 8387\ 6864)=2.775\ 56\times 
10^{-17}.$ Here, $\omega( 20.07\ 9666\ 0982\ 3604 
+ 7.513\ 2954\ 8387\ 6864\ )/2$ $=6.28319=2\pi.$
%

%
\begin{figure}[tbh]
\centering
\includegraphics[width=0.557\textwidth]{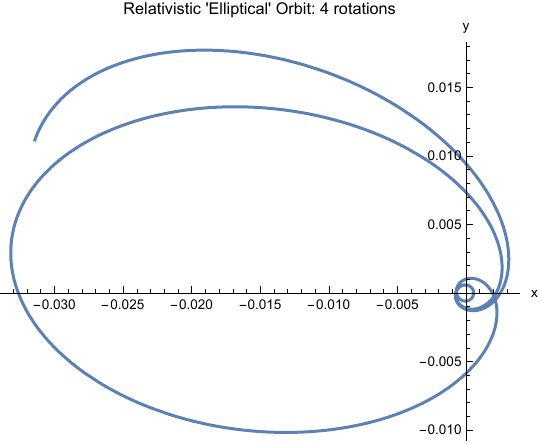}
\caption{Kepler's `Elliptical' Curve Counterclockwise Motion with Four Self-intersections in
Hypothetical Unbiennium Hydrogen-like Ion $Ube^{128+}\/.$ The `winding' number is four.}
\label{Figure5}
\end{figure}
%

%
%
\begin{figure}[tbh]
\centering
\includegraphics[width=0.557\textwidth]{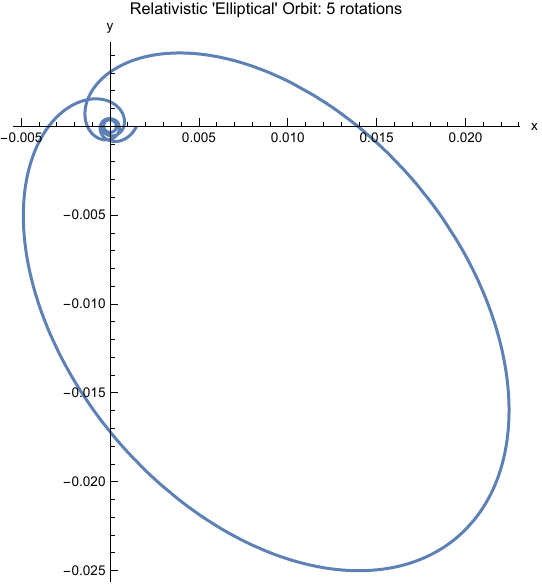}
\caption{Kepler's `Elliptical' Curve Counterclockwise Motion with Multiple Self-intersections in
Hypothetical Untribium Hydrogen-like Ion $Utb^{131+}\/.$ The `winding' number is five.}
\label{Figure6}
\end{figure}
%

%
%
\begin{figure}[tbh]
\centering
\includegraphics[width=0.557\textwidth]{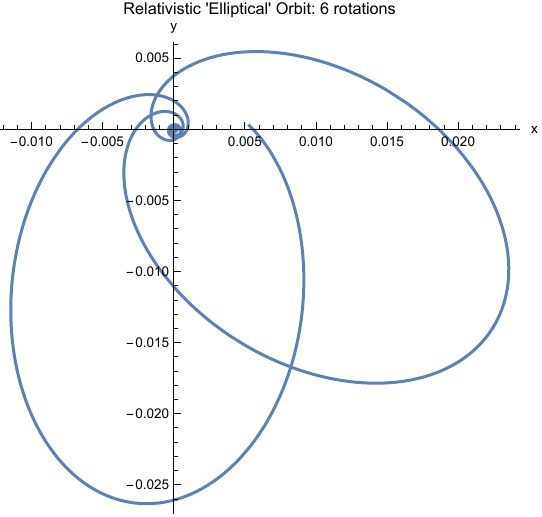}
\caption{Kepler's `Elliptical' Curve Counterclockwise Motion with Multiple Self-intersections in
Hypothetical Unbiennium Hydrogen-like Ion $Utp^{134+}\/.$ The `winding' number is six.}
\label{Figure7}
\end{figure}
%
%

\subsection{Unbiennium\/}  
In the above animation of the relativistic Kepler motion of an electron in a hypothetical
hydrogen-like Unbiennium ion, $Ube^{128+}$, when $Z=129$, the nucleus is situated in the
fixed focus at the origin. For this quantum state, we choose $n_{r}=n_{\theta}=1$ 
in the fine structure formula (\ref{BidSomEnd}). By (\ref{BidSom18})--(\ref{BidSom20}), one gets:
$\omega=0.337\ 408$, $\epsilon=0.967\ 653$, and $a=0.016\ 9559$, in Bohr's atomic units.

Once again, the perihelion and aphelion move along two concentric circles around the
nucleus with radii: $r_{\text{min}}=a(1-\epsilon)=0.000\ 5484\ 74$ and
$r_{\text{max}}=a(1+\epsilon)=0.033\ 3634$, respectively. In the animation,
the geometrical loci of the successive perihelia and aphelia, the outer and
inner envelopes of the orbit, are not shown for simplicity (Figure~\ref{Figure4}).
Rotation of the ellipse over one period is given by $\Delta\theta
=2\pi(1/\omega-1)=12.33\ 87.\/$ The `winding' number is four.
\subsection{Untribium\/} 
In the above animation of the relativistic Kepler motion of an electron in a hypothetical
hydrogen-like Unbiennium ion, $Utb^{131+}$, when $Z=132$, the nucleus is situated in the
fixed focus at the origin. For this quantum state, we choose $n_{r}=n_{\theta}=1$ 
in the fine structure formula (\ref{BidSomEnd}). By (\ref{BidSom18})--(\ref{BidSom20}), one gets:
$\omega=0.268\ 605$, $\epsilon=0.977\ 328$, and $a=0.015\ 3084$, in Bohr's atomic units.

The perihelion and aphelion move along two concentric circles around the
nucleus with radii: $r_{\text{min}}=a(1-\epsilon)=0.000\ 3470\ 77$ and
$r_{\text{max}}=a(1+\epsilon)=0.030\ 2698$, respectively. In the animation,
the geometrical loci of the successive perihelia and aphelia, the outer and
inner envelopes of the orbit, see Figure~\ref{Figure6}, are not shown for simplicity.
Rotation of the ellipse over one period is given by $\Delta\theta
=2\pi(1/\omega-1)=17.10\ 88.\/$ The `winding' number is five.
\subsection{Untripentium\/}   
In the above animation of the relativistic Kepler motion of an electron in a hypothetical
hydrogen-like Unbiennium ion, $Utp^{134+}$, when $Z=135$, the nucleus is situated in the
fixed focus at the origin. For this quantum state, we choose $n_{r}=n_{\theta}=1$ 
in the fine structure formula (\ref{BidSomEnd}). By (\ref{BidSom18})--(\ref{BidSom20}), one gets:
$\omega=0.171\ 738$, $\epsilon=0.989\ 201$, and $a=0.013\ 287$, in Bohr's atomic units.

The perihelion and aphelion move along two concentric circles around the
nucleus with radii: $r_{\text{min}}=a(1-\epsilon)=0.000\ 14349$ and
$r_{\text{max}}=a(1+\epsilon)=0.026\ 4305$, respectively. In the animation,
the geometrical loci of the successive perihelia and aphelia, the outer and
inner envelopes of the orbit, see Figure~\ref{Figure7}, are not shown for simplicity.
Rotation of the ellipse over one period is given by $\Delta\theta
=2\pi(1/\omega-1)=30.30\ 26.\/$ The `winding' number is six.

\vfill
\eject
\newpage
\noindent
\noindent
Orbit parameters under consideration are compared in the following tables for the reader's convenience:
%

%
\begin{tabular}
[c]{|l|l|l|l|}\hline
$Z$ & $U^{91+}$ Ion & $Cn^{111+}$ Ion & $Og^{117+}$ Ion\\\hline
$\omega$ & $0.741\ 135$ & $0.576\ 207$ & $0.508\ 457$\\\hline
$\epsilon$ & $0.904\ 882$ & $0.930\ 786$ & $0.941\ 479$\\\hline
$a/a_{0}$ & $0.035\ 3163$ & $0.024\ 9872$ & $0.022\ 2041$\\\hline
$r_{\text{min}}$ & $0.003\ 3592\ 09$ & $0.001\ 72947$ & $0.001\ 2994$\\\hline
$r_{\text{max}}$ & $0.067\ 2734\ 72$ & $0.048\ 2449$ & $0.043\ 1087$\\\hline
$\Delta\theta$ & $2.194\ 61$ & $4.621\ 2$ & $6.074\ 18$\\\hline
`Winding' $\#^{\prime}$s & $1$ & $2$ & $2$\\\hline
\end{tabular}
%

%
\noindent
%
\begin{tabular}
[c]{|l|l|l|l|l|}\hline
$Z$ & $Ubb^{121+}$ Ion & $Ube^{128+}$ Ion & $Utb^{131+}$ Ion & $Utp^{134+}$
Ion\\\hline
$\omega$ & $0.455\ 419$ & $0.337\ 408$ & $0.268\ 605$ & $0.171\ 738$\\\hline
$\epsilon$ & $0.949\ 782$ & $0.967\ 653$ & $0.977\ 328$ & $0.989\ 201$\\\hline
$a/a_{0}$ & $0.020\ 3534$ & $0.016\ 9559$ & $0.015\ 3084$ & $0.013\ 287$%
\\\hline
$r_{\text{min}}$ & $0.001\ 02211$ & $0.000\ 5484\ 74$ & $0.000\ 3470\ 77$ &
$0.000\ 14349$\\\hline
$r_{\text{max}}$ & $0.039\ 6847$ & $0.033\ 3634$ & $0.030\ 2698$ &
$0.026\ 4305$\\\hline
$\Delta\theta$ & $7.513\ 3$ & $12.33\ 87$ & $17.10\ 88$ & $30.30\ 26$\\\hline
`Winding' $\#^{\prime}$s & $3$ & $4$ & $5$ & $6$\\\hline
\end{tabular}

%
\noindent
Further animation details can be found in our complementary Mathematica
notebooks \cite{BarleySusMathTwo}. 

\section{Conclusion}

We have demonstrated that in a super strong static electric field, similar to the hypothetical
Oganesson hydrogen-like ion, there is a possibility of self-intersecting
trajectory, which is different from the standard classical orbits. 
A possible rotation of an incoming electron around the Oganesson nucleus during the scattering should also be investigated \cite{Somm1916}.  
It would be interesting to analyze a possibility of a similar `Oganesson-type' effect in the
theory of gravitation \cite{LaLif2} and to study non-standard consequences 
in quantum electrodynamics in the strong, \textit{U}, versus super strong, \textit{Og} and beyond, Coulomb fields 
\cite{AkhBer, Colloquium2019, FleKar1971, Gum2005, Gum2007, Kaygorodovetal2021, Mohretal1998, Oganessian2017, PomSmor1945, ShabQED, ShabQEDHI}. 
For example, in the creation of an electron-positron pair, the electron may be captured by the nucleus directly
from the internal orbit. Further nuclear reaction with one of Oganesson's proton may result in neutron and neutrino, 
thus contributing into the instability of this super heavy element.
In a hypothetical Unbibium, element with $Z=122\/,$ there is a possibility of a multiple self-intersecting 
trajectory, which should even more contribute into an orbit electron capture and 
the corresponding \textit{Ubb} transformation and beyond, when $Z\le135\/$.

Our examples may give an idea of a topological classification, -- by orbits `winding' numbers --, for the strength of Coulomb fields in the superheavy transuranium elements. 

\vfill
\eject
\newpage
{\scshape{Sommerfeld vs oganesson puzzles\/:}}
As well-known, the energy levels of a hydrogen-like system in Dirac's theory match, precisely, the Sommerfeld formula (\ref{BidSomEnd}) -- an outcome known as the \textquotedblleft Sommerfeld Puzzle\textquotedblright\ \cite{Biedenharn1983}, discussed further in \cite[pp.~426--429]{Eckert}. This \textquotedblleft puzzle\textquotedblright\ has recently been resolved \cite{SusPuzz}.
But why does the presence of an extra loop in the `old' Bohr-Sommerfeld, or semi-classical, trajectory of the electron in the transuranic heavy ions under consideration coincide with the parameters of artificially created elements on Earth, including Oganesson, which is located at the end of the Mendeleev periodic table?
 
\bigskip

\noindent \textbf{Acknowledgments} We are grateful to Professor~Ruben~Abagyan, Professor~Victor~V.~Dodonov, Dr.~Sergey~I.~Kryuchkov, 
Dr.~Eugene Stepanov, and Professor~Alexei Zhedanov for their insightful comments and help.

\end{document}